\documentclass[aps,prc,twoside,twocolumn,floatfix,superscriptaddress,amsmath,showpacs,amssymb]{revtex4}
\usepackage{amssymb}
\usepackage{bm}
\usepackage{graphicx}
\usepackage{psfrag}

\begin{document}

\def\buch{Institute for Nuclear Physics and Engineering, Bucharest, Romania}
\def\buda{KFKI Research Institute for Particle and Nuclear Physics, Budapest, Hungary}
\def\cler{Laboratoire de Physique Corpusculaire, IN2P3/CNRS, and Universit\'{e} Blaise Pascal, Clermont-Ferrand, France}
\def\sp{University of Split, Split, Croatia}
\def\darm{Gesellschaft f\"{u}r Schwerionenforschung, Darmstadt, Germany}
\def\dres{Institut f\"{u}r Strahlenphysik, Forschungszentrum Dresden-Rossendorf, Dresden, Germany} 
\def\heid{Physikalisches Institut der Universit\"{a}t Heidelberg, Heidelberg, Germany}
\def\mosc{Institute for Theoretical and Experimental Physics, Moscow, Russia}
\def\kurc{Kurchatov Institute, Moscow, Russia}
\def\seou{Korea University, Seoul, Korea}
\def\stra{Institut Pluridisciplinaire Hubert Curien and Universit\'{e} Louis Pasteur, Strasbourg, France}
\def\wars{Institute of Experimental Physics, Warsaw University, Warsaw, Poland}
\def\zagr{Ru{d\llap{\raise 1.22ex\hbox{\vrule height 0.09ex width 0.2em}}\rlap{\raise 1.22ex\hbox{\vrule height 0.09ex width 0.06em}}}er Bo\v{s}kovi\'{c} Institute, Zagreb, Croatia}
\def\lan{Institute of Modern Physics, Chinese Academy of Sciences, Lanzhou, China}
\def\ita{Laboratori Nazionali del Sud INFN, Catania, Italy}
\def\gat{Dept.~f\"{u}r Physik der Universit\"{a}t Muenchen, Garching, Germany}

\title[]{Isospin dependence of relative yields of $K^+$ and $K^0$ mesons at 1.528$A$ GeV}

\author{X. Lopez} \email{X.Lopez@gsi.de} \affiliation{\darm}
\author{Y.J. Kim} \email{Y.-J.KIM@gsi.de} \affiliation{\darm}
\author{N. Herrmann} \affiliation{\heid}
\author{A. Andronic} \affiliation{\darm}
\author{V. Barret} \affiliation{\cler}
\author{Z. Basrak} \affiliation{\zagr}
\author{N. Bastid} \affiliation{\cler}
\author{M.L. Benabderrahmane} \affiliation{\heid}
\author{R. \v{C}aplar} \affiliation{\zagr}
\author{E. Cordier} \affiliation{\heid}
\author{P. Crochet} \affiliation{\cler}
\author{P. Dupieux} \affiliation{\cler}
\author{M. D\v{z}elalija} \affiliation{\sp}
\author{Z. Fodor} \affiliation{\buda}
\author{I. Ga\v{s}pari\'c} \affiliation{\zagr}
\author{Y. Grishkin} \affiliation{\mosc}
\author{O.N. Hartmann} \affiliation{\darm}
\author{K.D. Hildenbrand} \affiliation{\darm}
\author{B. Hong} \affiliation{\seou}
\author{T.I. Kang} \affiliation{\seou}
\author{J. Kecskemeti} \affiliation{\buda}
\author{M. Kirejczyk} \affiliation{\wars}
\author{M. Ki\v{s}} \affiliation{\darm} \affiliation{\zagr}
\author{P. Koczon} \affiliation{\darm}
\author{M. Korolija} \affiliation{\zagr}
\author{R. Kotte} \affiliation{\dres}
\author{A. Lebedev} \affiliation{\mosc}
\author{Y. Leifels} \affiliation{\darm}
\author{M. Merschmeyer} \affiliation{\heid}
\author{W. Neubert} \affiliation{\dres}
\author{D. Pelte} \affiliation{\heid}
\author{M. Petrovici} \affiliation{\buch}
\author{F. Rami} \affiliation{\stra}
\author{W. Reisdorf} \affiliation{\darm}
\author{M.S. Ryu} \affiliation{\seou}
\author{A. Sch\"{u}ttauf} \affiliation{\darm}
\author{Z. Seres} \affiliation{\buda}
\author{B. Sikora} \affiliation{\wars}
\author{K.S. Sim} \affiliation{\seou}
\author{V. Simion} \affiliation{\buch}
\author{K. Siwek-Wilczy\'{n}ska} \affiliation{\wars}
\author{V. Smolyankin} \affiliation{\mosc}
\author{G. Stoicea} \affiliation{\buch}
\author{Z. Tyminski} \affiliation{\wars}
\author{P. Wagner} \affiliation{\stra}
\author{K. Wi\'{s}niewski} \affiliation{\wars}
\author{D. Wohlfarth} \affiliation{\dres}
\author{Z.G. Xiao} \affiliation{\lan}
\author{I. Yushmanov} \affiliation{\kurc}
\author{X.Y. Zhang} \affiliation{\lan}
\author{A. Zhilin} \affiliation{\mosc}

\collaboration{FOPI Collaboration} 
\noaffiliation

\author{G. Ferini } \affiliation{\ita}
\author{T. Gaitanos} \affiliation{\gat}

\date{\today}

\begin{abstract}
Results on $K^+$ and $K^0$ meson production in $^{96}_{44}$Ru + $^{96}_{44}$Ru and $^{96}_{40}$Zr + $^{96}_{40}$Zr collisions at a beam kinetic energy of 1.528$A$ GeV, measured with the FOPI detector at GSI-Darmstadt, are investigated as a possible probe of isospin effects in high density nuclear matter. The measured double ratio ($K^+/K^0$)$_{Ru}$/($K^+/K^0$)$_{Zr}$ is compared to the predictions of a thermal model and a Relativistic Mean Field transport model using two different collision scenarios and under different assumptions on the stiffness of the symmetry energy. We find a good agreement with the thermal model prediction and the assumption of a soft symmetry energy for infinite nuclear matter while more realistic transport simulations of the collisions show a similar agreement with the data but also exhibit a reduced sensitivity to the symmetry term.
\end{abstract}


\maketitle

\section{\label{sec:intro}Introduction}

One of the main motivation to study relativistic heavy ion collisions at intermediate energies is to obtain information on the equation-of-state (EoS) for nuclear matter at extreme conditions of pressure and density \cite{dani1}. Within the high beam kinetic energy range of SIS (1-2$A$ GeV), the system could reach three times the normal nuclear density and a temperature of about 60 MeV \cite{dani1,friman}. Presently there are strong claims for a soft EoS based on the system size dependence of the $K^+$ yields \cite{sturm,fuchs,hart,fuchs2} and on the azimuthal dependence of the mean kinetic energy of Z=1 particles \cite{gabi}. However, when considering the excitation function of differential directed \cite{anton} and elliptic flow \cite{anton1} of charged particles in the incident energy range from 0.1 to 1.5$A$ GeV, the  stiffness of the EoS appears less consistent, despite earlier preference for a soft EoS \cite{dani1} from similar measurements.

The EoS of asymmetric nuclear matter expressed as the energy per nucleon as a function of baryonic density $\rho_B$ and isospin asymmetry $\alpha$ is usually described as follows \cite{li1,liu}:
\begin{subequations}
\begin{equation}
E(\rho_B,\alpha)=E(\rho_B,\alpha=0)+E_{sym}(\rho_B)\alpha^2+O(\alpha^4),
\label{subeq:1}
\end{equation}
\begin{equation}
E_{sym}(\rho_B)=\frac{1}{2}\frac{\partial^2E(\rho_B,\alpha)}{\partial\alpha^2}\vert_{\alpha=0},~~~~\alpha=\frac{N-Z}{N+Z}.
\label{subeq:2}
\end{equation}
\end{subequations}
$E_{sym}$ is the nuclear symmetry energy per nucleon and $\alpha$ the asymmetry parameter.

Predictions for the density dependence of $E_{sym}$ based on various many-body theories diverge widely \cite{li1,liu,li2,gait,gait1}. Even the sign of the symmetry energy at baryonic densities above three times the normal nuclear density is still uncertain \cite{li4}.
The study of the isospin dependent part of the EoS is important to understand astrophysical processes like the structure of neutron stars and supernovae explosions \cite{be1,klahn} but also the structure of nuclei far off the valley of stability \cite{br1,st1}.
Recently, the momentum dependence of the symmetry potential was implemented in transport models in order to describe its effect in heavy-ion collisions more precisely \cite{li3,li5}.

Several observables, for example $n/p$ and $\pi^-/\pi^+$ yield ratios, are shown to be sensitive probes of the high density behavior of the nuclear symmetry energy within transport model calculations \cite{li1,li2,gait,gait1}.
According to recent theoretical results \cite{ditoro1,gait}, the $K^+/K^0$ yield could be very sensitive to the isospin contribution to the EoS. For beam energies near kaon production threshold, kaon yields should carry information about the dense and hot system created during the collision \cite{sturm,fuchs,hart28,li28,hart,aichel}.

In this work, we investigate the behavior of the symmetry energy of the nuclear matter by comparing the relative production yields of $K^+$ and $K^0$ to theoretical predictions \cite{anton2,ditoro1,theopri}. The study of the double ratio ($K^+/K^0$)$_{Ru}$/($K^+/K^0$)$_{Zr}$ permits to isolate the isospin contribution on kaon production in systems ($^{96}_{44}$Ru and $^{96}_{40}$Zr) which have the same atomic mass but a different number of neutrons. The FOPI collaboration has performed an experiment to maximize differences in the asymmetry (Eq.~\ref{subeq:2}) with $^{96}_{44}$Ru + $^{96}_{44}$Ru and $^{96}_{40}$Zr + $^{96}_{40}$Zr systems corresponding to asymmetry parameters $\alpha$ of 0.083 and 0.167, respectively. The present results are obtained at 1.528$A$ GeV beam kinetic energy ($\sqrt{s_{NN}}=2.53$ GeV), slightly below the threshold energy of $K^+$ and $K^0$ production. Isospin effects measured at 400$A$ MeV for the same colliding nuclei have been reported in \cite{h1,ra1}.

The paper is structured as follows. Section II consists of a short description of the apparatus and the event characterization. The identification method of $K^+$ and the reconstruction method of $K^0_S$ are described in section III. In section IV the experimental $K^+/K^0$ ratio measured in Ru and Zr systems is compared to theoretical model predictions. Finally, we summarize and give an outlook in section V.

\section{\label{sec:exprm}Experimental set up}

The experiment has been performed with the FOPI detector at the Heavy-Ion Synchrotron SIS of GSI-Darmstadt by using $^{96}_{44}$Ru and $^{96}_{40}$Zr beams of kinetic energy of 1.528$A$ GeV on enriched targets ($>$95\%) of $^{96}_{44}$Ru and $^{96}_{40}$Zr.
The beam intensity was $3 \times 10^4$ ions/s, and the target thicknesses were 431 and 380 mg/cm$^2$ for Ru and Zr, which is equivalent to an interaction probability of 1\%. More details on the experiment can be found in \cite{h1,ra1}.

The FOPI detector is an azimuthally symmetric apparatus made of several sub-detectors which provide charge and mass determination over nearly the full solid angle. The central part of the detector, placed in a super-conducting solenoid, consists of a Central Drift Chamber (CDC) surrounded by plastic scintillators (Barrel). The forward part is composed of a wall of plastic scintillators (PLAWA and ZDD) and another drift chamber (Helitron) placed inside the super-conducting solenoid.

For the present analysis, the PLAWA, the Barrel and the CDC were used. Particles measured in the CDC are identified by their mass, which is determined by the correlation between magnetic rigidity and specific energy loss. The Barrel and the PLAWA provide charge identification of the reaction products, combining time of flight (ToF) and specific energy loss information. More details on the configuration and performances of the different components of the FOPI apparatus can be found in \cite{gob,rit1,andro1}.

The events are selected by their centrality according to the charged particle multiplicities measured in the PLAWA and in the CDC. The results presented in the following are obtained for the most central 260 mb cross section, which corresponds to 7$\%$ of the total geometrical cross section.

\section{\label{sec:rec}Identification and reconstruction of kaons}

\subsection{Identification of $K^+$}

$K^+$ mesons are identified by correlating their momentum determined with the CDC and their velocity measured with the ToF Barrel. In addition, a cut on the laboratory momentum ($p_{lab} < 0.5$ GeV/$c$) is applied in order to reduce the contamination due to fast $\pi^+$ mesons and protons. More details on the $K^+$ identification are provided in \cite{chechek,crochet}. The phase space occupancy of $K^+$ in both systems (Ru + Ru and Zr + Zr) corresponds to a covering of the laboratory polar angle $\theta_L$ from $39^\circ$ to $130^\circ$. The efficiency of $K^+$ identification is the same for the two systems and the fraction of measured kaons is about 12\% of the total number produced in 4$\pi$.  
\begin{figure}[!th]
\vspace{-0.7cm}\includegraphics[width=8cm]{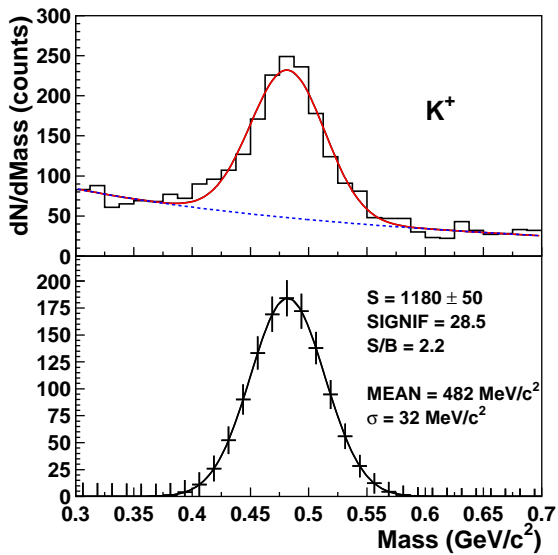}
\caption{\label{fig:minvk+} Mass spectra of $K^+$, the solid and dashed lines denote the signal and background, respectively (upper panel). The lower panel shows the signal after background subtraction. The following characteristics of the signal are reported: number of counts in the signal (S), significance (SIGNIF) and signal-to-background ratio (S/B) at $\pm 2\sigma$. The fitted parameters of the Gaussian function as the mean value of the mass peak (MEAN) and the Gaussian width ($\sigma$) are also mentioned.}
\end{figure}

An example of the $K^+$ mass spectrum is shown in Fig.~\ref{fig:minvk+} for Zr + Zr collisions. The distribution was fitted with exponential and Gaussian functions in the mass region between 0.3 and 0.7 GeV/$c^2$. The background is estimated with the exponential function while the signal is described by the Gaussian function.
The signal after background subtraction is plotted in the lower panel of Fig.~\ref{fig:minvk+}. Within an interval of $\pm 2\sigma$, about 1180 $K^+$ are identified for the applied set of cuts in the analysis. The signal-to-background ratio is about 2.2 and the mass width ($\sigma$) of 32 MeV/$c^2$ is dominated by the detector resolution. The statistical error of the signal is evaluated with the covariant matrix by using the error on Gaussian parameters.

The systematic error of this measurement is determined by applying different cuts on sensitive variables used to identify $K^+$ such as the matching quality between the CDC and the Barrel. Also, another description of the background by using a polynomial function was used. After the determination of the number of counts in the mass spectra, a normalization with respect to the number of events was applied in order to directly compare the two colliding systems. Finally, the ratio of the production yields of $K^+$ measured in the two systems is:
\begin{equation}
\label{rat1}
K^+_{Ru}/K^+_{Zr} = 1.06\pm 0.07(\mathrm{stat.}) \pm 0.09(\mathrm{syst.})
\end{equation}

\subsection{Reconstruction of $K^0$}

Due to $K^0-\overline{K^0}$ mixing, $K^0_L$ and $K^0_S$ are produced in equal amounts. As a consequence of their lifetime and decay channels, the $K^0_L$ can not be measured with the FOPI detector whereas the $K^0_S$ can be reconstructed via the weak decay into $\pi^-$ and $\pi^+$ (branching ratio = 68.6$\%$, $c\tau = 2.68$ cm). The lifetime of $K^0_S$ is long enough to permit the reconstruction of its vertex decay in the CDC. Therefore it is possible to significantly improve the signal-to-background ratio by rejecting particles coming from the primary vertex. The phase space occupancy of $K^0_S$ in both systems (Ru + Ru and Zr + Zr) corresponds to a covering of the laboratory polar angle $\theta_L$ from $32^\circ$ to $140^\circ$.

The $K^0_S$ are reconstructed from identified pions by applied a cut on the laboratory momentum ($p_{lab} \ge 0.08$ GeV/$c$) in order to reject pions which are spiraling in the CDC. The distance of closest approach (DCA) between tracks and the primary vertex in the transverse plane (DCA $>$ 0.7 cm) is used to enhance the fraction of particles coming from a secondary vertex in the data sample. The efficiency of $K^0_S$ reconstruction is the same for the two systems. With the above mentioned cuts, we reconstruct 8\% of all $K^0_S$ produced.

\begin{figure}[!th]
\vspace{-0.5cm}\includegraphics[width=8cm]{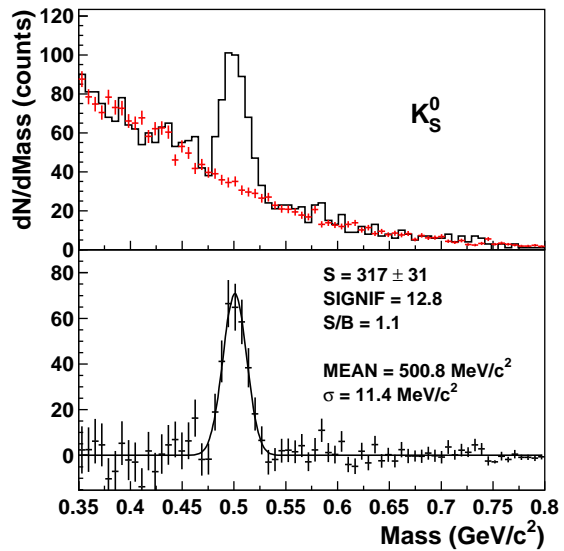}
\caption{\label{fig:minv} Invariant mass spectra of $\pi^-\pi^+$ pairs. The solid line and crosses denote the combinatorics and the scaled mixed-event background, respectively (upper panel). The lower panel shows the signal after background subtraction.}
\end{figure}

The invariant mass of the particle pair is calculated from the four-momenta of the pions at the intersection point. The reconstruction of $K^0_S$ measured for the Zr + Zr system is shown in Fig.~\ref{fig:minv}. 

The combinatorial background is obtained by using the event-mixing method \cite{berger}. For that purpose, the two decay particles are taken from two different events which have the same centrality selection. In addition, the two events are rotated into the reaction plane in order to have the same reference system for both particles. This latter is estimated event by event according to the standard transverse momentum procedure \cite{dan85}. The mixed-event background is normalized in order to fit the combinatorics in the invariant mass range 0.35-0.45 GeV/$c^2$. The shape of the resulting mixed-event background describes the combinatorial background and is indicated by crosses in Fig.~\ref{fig:minv} (upper panel). The vertical bars of the crosses correspond to the statistical errors. After background subtraction (Fig.~\ref{fig:minv}, lower panel), the remaining peak in the mass spectra is fitted with a Gaussian function. Within an interval of $\pm 2\sigma$, about 320 $K^0_S$ are found for the applied set of cuts in the analysis. The signal-to-background ratio is about 1.1 and the width ($\sigma=11.4$ MeV/$c^2$) of the distribution reflects the detector~resolution.

The statistical error is determined by using the same method as for $K^+$. In order to estimate the systematic error of this measurement we apply different sets of cuts by varying the selection criteria of the relevant quantities discussed in the beginning of this section. Finally, we obtain the following ratio of the production yields of $K^0$ for Ru + Ru and Zr + Zr systems:
\begin{equation}
\label{rat2}
K^0_{Ru}/K^0_{Zr} = 0.94\pm 0.12(\mathrm{stat.})\pm 0.06 (\mathrm{syst.}).
\end{equation}

\section{\label{sec:reslt}Model comparison}

The experimental double ratio obtained~in~this~analysis~is:
\begin{subequations}
\begin{equation}
\frac{K^+_{Ru}/K^+_{Zr}}{K^0_{Ru}/K^0_{Zr}}=\frac{(K^+/K^0)_{Ru}}{(K^+/K^0)_{Zr}},
\label{subeq:3}
\end{equation}
\begin{equation}
\frac{(K^+/K^0)_{Ru}}{(K^+/K^0)_{Zr}}=1.13\pm 0.16(\mathrm{stat.})\pm 0.12 (\mathrm{syst.}). 
\label{subeq:4}
\end{equation}
\end{subequations}
Note that by using this double ratio, the effect of different phase space occupancies for $K^0$ and $K^+$ mesons are cancelled out (Eq.~\ref{subeq:3}).
\begin{figure}[!th]
\vspace{-0.5cm}\includegraphics[width=8cm]{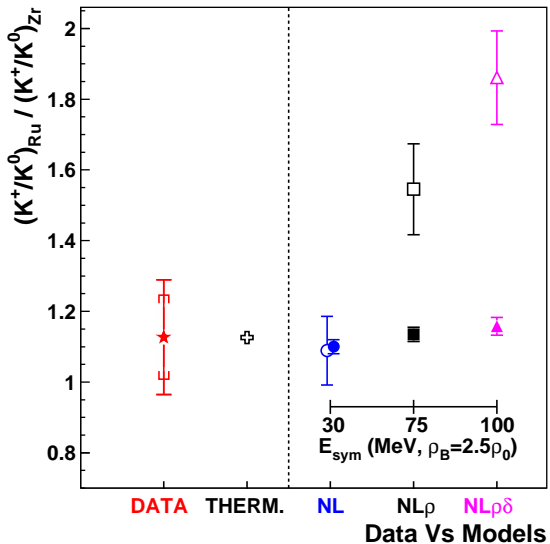}
\caption{\label{fig:comp} Experimental ratio ($K^+/K^0$)$_{Ru}$/($K^+/K^0$)$_{Zr}$ (star) and theoretical predictions of the thermal model (cross) and the transport model with 3 different assumptions on the symmetry energy: NL (circles), NL$\rho$ (squares) and NL$\rho \delta$ (triangles). The INM and HIC calculations are represented by open and full symbols, respectively (see text for more details). The statistic and systematic errors are represented by vertical bars and brackets, respectively.}
\end{figure}

The experimental result is compared to a thermal model prediction \cite{anton2} (left part of Fig.~\ref{fig:comp}) using a temperature of $T=52$ MeV and a baryonic chemical potential of $\mu_{B}=756$ MeV. It is important to stress that the thermal model assumes full stopping and mixing of the two nuclei and ensures isospin conservation via a chemical potential ($\mu_{I_3}$ = -7.1 MeV and -13.5 MeV for Ru + Ru and Zr + Zr systems, respectively). Also included in Fig.~\ref{fig:comp} (right part) are two sets of calculations based on Relativistic Mean Field theory (RMF) \cite{ditoro1,gait,gait1,theopri}. One assumes infinite nuclear matter (INM, open symbols) using a temperature of $T=60$ MeV and a baryonic density of $\rho_B=2.5 \rho_0$ \cite{ditoro1} and the second describes a finite system as realized in heavy ion collisions (HIC, full symbols) \cite{theopri}. Three different assumptions for the symmetry energy are used in the transport model calculations. The reference ratio obtained from a non-linear Walecka model \cite{gait1} (NL, illustrated by circles in Fig.~\ref{fig:comp}) represents the prediction with a soft symmetry energy, only the kinetic contribution is taken into account. Assumptions labelled NL$\rho$ (squares) and NL$\rho\delta$ (triangles) show the expectation of the model including the vector part ($\rho$) and both vector part plus scalar part ($\rho+\delta$) of the isovector mean field, which correspond to a gradually stiffer symmetry energy \cite{ditoro1,gait,gait1}. The corresponding value of the symmetry energy is reported on Fig.~\ref{fig:comp} (right part) for the INM case. HIC calculations have been performed explicitly for $^{96}_{44}$Ru + $^{96}_{44}$Ru and $^{96}_{40}$Zr + $^{96}_{40}$Zr systems, while INM calculations are taken from \cite{ditoro1}.

The prediction of the thermal model is in perfect agreement with the data. It is worth to point out that the thermal model calculations reproduce very well the measured values of the single ratios of $K^+$ and $K^0$ in Ru and Zr systems (Eq.~\ref{rat1} and Eq.~\ref{rat2}, respectively). In the INM calculations the system is in equilibrium (chemical, thermodynamical, isospin) and with a soft symmetry energy (NL, no isospin dependent collision term) the model prediction can be directly compared to the thermal model calculation. As expected, both predictions are in agreement. Furthermore, the transport model permits to vary the stiffness of the symmetry energy as it is shown in Fig.~\ref{fig:comp} (right part). An increase of the ($K^+/K^0$)$_{Ru}$/($K^+/K^0$)$_{Zr}$ ratio as a function of the symmetry energy is observed for both the INM and the HIC calculations. With the INM scenario, a large enhancement ($\sim$70\%) is predicted when going from a soft (NL) to a stiff (NL$\rho\delta$) EoS. The comparison with the experimental point shows that our result is in favor of a soft symmetry energy. The isospin dependence, however, is dramatically reduced, when treating the EoS within transport model calculations. Taking into account the experimental errors, the measurement does not permit to distinguish between the three different assumptions introduced in the HIC scenario. The sensitivity on isospin dependence of this observable is reduced to 5\% between the two most extreme cases (NL and NL$\rho \delta$). In contrast to the static INM calculations, in the HIC scenario two important dynamical effects occur:~fast neutron emission (mean field effect) and transformation of neutron into proton in inelastic channels (no-chemical equilibrium). This appears to be the major reason for the reduction of the isospin effect on kaons ratios. Finally, for kaons measured at threshold energy from systems with asymmetry parameters differing by a factor two, the isovector contribution in nucleon-nucleon interaction influences very marginally the relative yields of $K^+$ and $K^0$. Recent theoretical calculations \cite{toroend} suggest that a measurement of the double ratio at lower energies and with systems exhibiting a large difference in their asymmetry might increase the sensitivity on the isospin dependence of kaon production. On the experimental side, measuring a sample of kaons with sufficiently high statistics at sub-threshold energies will be a considerable challenge.

\section{\label{sec:con}Conclusions}
We have reported the first measurement of the relative production yield of $K^+$ and $K^0$ mesons in Ru + Ru and Zr + Zr systems at a beam kinetic energy of 1.528$A$ GeV which is close to the kaon production threshold.

The experimental ($K^+/K^0$)$_{Ru}$/($K^+/K^0$)$_{Zr}$ ratio is compared to a thermal model and a model based on the Relativistic Mean Field theory using three different assumptions on the symmetry energy and two different collision scenarios. The measurement is in good agreement with the predictions of the thermal model and of the RMF model for infinite nuclear matter with a soft symmetry energy. The RMF model for a collision system (HIC) shows a similar agreement with the data but the sensitivity to the symmetry term decreases by more than one order of magnitude. Within the model described above, the isovector contribution of nucleon-nucleon interaction influences weakly the relative yields of kaons measured at threshold energy and for systems differing in the asymmetry term by a factor two. A good understanding of this phenomenon together with the present measurement will have a strong impact when designing new experiments aiming at the isospin dependence of the nuclear equation-of-state.

\begin{acknowledgments}
We are grateful to C. Fuchs, H.H. Wolter and M. Di Toro for providing us model calculations and for intensive discussions. This work was partly supported by the German BMBF under Contacts No.~06HD953, by the Korea Science and Engineering Foundation (KOSEF) under grant No. F01-2004-000-10002-0 and agreement between GSI and IN2P3/CEA.
\end{acknowledgments}

\end{document}